\documentclass[preprint] {article}
\usepackage{amsmath}
\usepackage{epsfig}
\usepackage{amsfonts}
\usepackage{amssymb}
\usepackage{geometry}
\usepackage{graphicx}
\allowdisplaybreaks
\newcommand {\bp}{\begin{pmatrix}}
\newcommand {\ep}{\end{pmatrix}}
\newcommand{\be}{\begin{equation}} \newcommand{\ee}{\end{equation}}
\newcommand{\bea}{\begin{eqnarray}}\newcommand{\eea}{\end{eqnarray}}

\begin{document}
\title{$\cal{PT}$-symmetric rational Calogero model with balanced loss and gain
}

\author{Debdeep Sinha\footnote{{\bf email:}  debdeepsinha.rs@visva-bharati.ac.in} 
\ and Pijush K. Ghosh \footnote {{\bf email:}
pijushkanti.ghosh@visva-bharati.ac.in}}
\date{Department of Physics, Siksha-Bhavana, \\ 
Visva-Bharati University, \\
Santiniketan, PIN 731 235, India.}
\maketitle

\begin{abstract}

A two body rational Calogero model with balanced loss and gain
 is investigated. The system yields a Hamiltonian which is symmetric 
under the combined operation of parity ($\cal{P}$) and time reversal ($\cal{T}$) symmetry. 
It is shown that the system is integrable and 
exact, stable classical solutions are obtained for particular ranges of the parameters.
The corresponding quantum system admits  bound state solutions for exactly the 
same ranges of the parameters for which the classical solutions are stable.
The eigen spectra of the system is presented with a discussion on the
 normalization of the wave functions in 
proper Stokes wedges. Finally, the Calogero model with balanced loss and gain
  is studied classically, when the pair-wise harmonic interaction term is replaced by a
common confining harmonic potential. The system admits stable solutions
for particular ranges of the parameters. However, the integrability and/or exact solvability of the
system is obscure due to the presence of the loss and gain terms.
The perturbative solutions are obtained and are compared with the  numerical results.

\end{abstract}

{\bf keywords:} Calogero model, $\cal{PT}$ symmetry, Balanced loss and gain.

\tableofcontents
\vspace{0.3in}

\section{Introduction}

The damped harmonic oscillator with a friction term linear in velocity is not 
a Hamiltonian system. In order to make the system Hamiltonian one needs 
to introduce a time reversed version of the original oscillator
which may be considered as a thermal bath\cite{bateman}.
A system consisting of these two oscillators considered together yields a 
$\cal{PT}$ symmetric Hamiltonian and the total energy is conserved. 
The Hamiltonian formulation necessarily implies that loss and gain are equally 
balanced. The quantization of this kind of coupled oscillators having balanced loss and
gain is also discussed in the literature\cite{cee1}\cite{cee2}\cite{cee3}\cite{cee4}. Neither classically stable solutions nor quantum bound states
can be obtained for this system. However, the situation changes significantly if these two oscillators
are coupled through interactions which are $\cal{PT}$ symmetric. An investigation in this regard
has been carried out recently\cite{ben}\cite{pra}, where a system of coupled oscillators having balanced loss
 and gain is considered. Both classically stable solutions as well as quantum bound states are obtained
within the unbroken $\cal{PT}$ symmetric region. Further, this system exhibits $\cal{PT}$ symmetric
phase transitions which occur at the same ranges of coupling parameter for the classical as well as quantum cases.
This mathematical model is motivated by an experiment performed on  two coupled $\cal{PT}$-symmetric 
whispering-gallery-mode optical resonators\cite{bpeng}. The results of this experiment are well explained 
by the model considered in Ref.\cite{ben}.

The oscillator systems having balanced loss and gain with 
different types of couplings are studied extensively in the literature. For example,
a pair of mutually coupled active LCR circuits, one with amplification and the other with
equivalent attenuation, is used to realize the $\cal{PT}$-symmetric phase transition\cite{JC}.
A chain of linearly coupled oscillators with its continuum limit is considered in\cite{ben1}.
 Further, $\cal{PT}$-symmetric dimer of coupled nonlinear oscillators with cubic nonlinearities is considered in \cite{khare}.
The system is not amenable to Hamiltonian formulation. Subsequently, a Hamiltonian system of nonlinear
 oscillators with balanced loss and gain is considered in\cite{freda}.
The small amplitude oscillations in this system are shown to be governed by a $\cal{PT}$ symmetric
nonlinear Schrodinger dimer. The common feature of all of these systems is that the unbroken
 $\cal{PT}$-symmetric regime content classically stable solutions.

 The Calogero model  \cite{calo}\cite{calo1}\cite{calo2}\cite{sut}\cite{sut1}\cite{sut2} is an exactly solvable model in one dimension
where each particle interacts with all other via a long range inverse square potential.
There are reviews\cite{ob}\cite{ob1}\cite{poly}\cite{pkg1} on  
the topic discussing various aspects of this model.
The Calogero type of systems have its influences in diverse branches of physics such as in
exclusion statistics\cite{exst}, quantum chaos\cite{quch},\cite{quch1}, spin chains\cite{hal}\cite{hal1} algebraic and integrable
structure\cite{integ}\cite{integ1}, self-adjoint extensions\cite{pkg2}\cite{pkg21}\cite{pkg22}\cite{pkg23},
 collective field formulation of 
many-particle systems\cite{cfmp}\cite{cfmp1} etc.. 
Therefore, the effect of Calogero type of potential in case of coupled oscillator system
having balanced loss and gain is an obvious curiosity. 
One of our main objectives in this work is to examine how the integrable properties of the
Calogero model get modified due to the presence of balanced loss and gain terms.  
 Another motivation of our study is to investigate 
$\cal{PT}$ symmetric phase transition in the presence of specified type of nonlinear interaction
governed by inverse square potential.
 Finally, it is expected that 
the additional inverse square interaction term may be realized in the context of 
whispering-gallery-mode optical resonators.

In this article, we investigate a two body rational Calogero model with balanced loss and gain.
 The system yields a Hamiltonian which is  $\cal{PT}$ symmetric. 
We obtain exact stable classical solutions  for the particular ranges of the parameters for
which the $\cal{PT}$ symmetry remains unbroken.
A quantization of this classical model is carried out.
This quantized version admits  bound state solutions for exactly the 
same ranges of the parameters for which the classical solutions are stable.
The eigen spectra and eigen states of the system are presented.
The eigen functions are not normalizable along the real line. We define the
proper Stoke wedges and discuss the normalization of the ground state wave function in 
this Stokes wedges. Finally, the Calogero model with balanced loss and gain
  is studied classically, when the pair-wise harmonic interaction term is replaced by a
common confining harmonic potential. In this case the system admits stable solutions
in the unbroken  $\cal{PT}$ symmetric regime. However, the exact solvability of the
system is obscure due to the presence of the loss and gain terms.
 For this case we obtained perturbative solutions in the unbroken ($\cal{PT}$) symmetric regime.
This perturbative results are compared with the  exact numerical calculations.

The plan of this paper is as follows. In the next section we introduce the main model. In section 2.1, we discuss the classical results
for a two body rational Calogero model with balanced loss and gain. The exact stable solution for this system is obtained. 
In section 2.2 , the classical perturbative solutions for the Calogero model with balanced loss and gain are discussed
 when the pair-wise harmonic interaction term is replaced by a
common confining harmonic potential. These results are compared with the numerical calculations.
 In section 3, the quantum
case for two body rational Calogero model with balanced loss and gain is considered.
 In the last section we make a summary and discuss the results.
 
\section{Classical model}

 The model we consider has the following Lagrangian: 

\be
L=\dot{x}\dot{y}+\gamma(x\dot{y}-y\dot{x})-\omega^2 xy -\frac{\epsilon}{2}(x^2+ y^2)-\frac{g}{2(x-y)^2},
\label{lag0}
\ee

\noindent where the dot over the variables denotes derivative with respect to time. The first four terms describe a pair of two oscillators, 
having common frequency $\omega$, with balanced loss and gain
and are coupled linearly via coupling parameter $\epsilon$. 
 The last term is the reminiscent
of the two-body Calogero potential describing a system of two particles interacting with each other via long range inverse square potential.
Thus the Lagrangian of Eq. (\ref{lag0}) describes a dissipative harmonic oscillator system along with its time reversed version 
 interacting with each other via a two-body inverse square potential plus a linear interacting term. The whole system yields a 
Hamiltonian 

\be
H=P_xP_y+\gamma(yP_y-xP_x)+(\omega^2-\gamma^2)xy+\frac{g}{2(x-y)^2}+\frac{\epsilon}{2}(x^2+y^2)
\label{hamil0}
\ee
where
\bea
P_x=\dot{y}-\gamma y,
\label{px}\ \ \ \ 
P_y=\dot{x}+\gamma x
\label{py}
\eea
\noindent are respectively the momenta conjugate to the $x$ and $y$ variables.
 The total  energy of the system is conserved. 

The following Eqs. of motions may be obtained either from the Lagrangian (\ref{lag0}) or from
 the Hamiltonian (\ref{hamil0}): 
\bea
\ddot{x}+2\gamma \dot{x}+(\omega^2x+\epsilon y)+\frac{g}{(x-y)^3}=0,
\nonumber\\
\ddot{y}-2\gamma \dot{y}+(\omega^2y+\epsilon x)-\frac{g}{(x-y)^3}=0.
\label{em0}
\eea
\noindent The rational Calogero model with the balanced loss and gain is obtained in the limit
 $\epsilon=-\omega^2$ for which two particles interact with each other via pair-wise harmonic
plus inverse square interaction. The limit $\epsilon=0$ corresponds to a system with balanced loss 
and gain where two particles are confined in a common harmonic potential and interacting with each
other through inverse square potential. The $\epsilon=0$ case may be considered as Sutherland model
\cite{sut} in the presence of balanced loss and gain terms. One of the purposes of this article is to
investigate whether or not the quantum and classical integrability of the Calogero-Sutherland model
is preserved after the inclusion of balanced loss and gain terms. The classical equations of motion
are highly nonlinear for $g\ne 0$. It is also under the purview of the present article to investigate 
$\cal{PT}$ symmetric phase transition in the presence of nonlinear interaction. Finally, the system
with $g=0$ is experimentally realized\cite{bpeng}\cite{ben}. It is expected that the additional inverse square
interaction term may be realized in the context of whispering-gallery-mode optical resonators.

\noindent It may be noted that the Hamiltonian (\ref{hamil0}) is $\cal{PT}$ symmetric where the action of 
parity $\cal{P}$ is to interchange the gain and loss oscillators:

\bea
{\cal{P}}:   x{ \rightarrow} -y,\ \ {\cal{P}}:   y{ \rightarrow} -x,\ \ {\cal{P}}:   P_x{ \rightarrow} -P_y,\ \ {\cal{P}}:  P_ y{ \rightarrow} -P_x,
\label{p}
\eea
and the action of time reversal $\cal{T}$ is to change the sign of the momenta:
 
\bea
{\cal{T}}:   x{ \rightarrow} x,\ \ {\cal{T}}:   y{ \rightarrow} y,\ \ {\cal{T}}:   P_x{ \rightarrow} -P_x,\ \ {\cal{T}}:  P_ y{ \rightarrow}- P_y.
\eea
It may be noted that if a real, time independent potential $V(x,y)$ is added to the Lagrangian (\ref{lag0}), then it remains $\cal{PT}$
symmetric provided $V(x,y)=V(-x,-y)$.

\noindent It will be convenient to cast the Eqs. of motion in the following new coordinates:

\bea
z_{1}=x+y, \ \ \ \ z_{2}= x-y.
\label{normal}
\eea

\noindent In this coordinates the Eqs. (\ref{em0}) take the following form:

\bea
\ddot{z}_{1}+(\omega^2 +\epsilon) z_1+2 \gamma\dot{z}_{2}=0,
\nonumber\\
\ddot{z}_{2}+(\omega^2-\epsilon)z_2+2 \gamma\dot{z}_{1}+\frac{2g}{z^3_2}=0.
\label{noreom}
\eea

\noindent It is interesting to see that under parity ($\cal{P}$), $z_1$ and $z_2$ transform as

\bea
{\cal{P}}:   z_1{ \rightarrow} -z_1,\ \ {\cal{P}}:   z_2{ \rightarrow} z_2.
\eea

\noindent Thus the parity operator has its usual meaning of
spatial reflection in the new coordinate system.

\subsection{Two body Calogero model with balanced loss and gain }

In the coordinates $(z_1,z_2)$, the Lagrangian of Eq. (\ref{lag0}) takes the following form for $\epsilon=-\omega^2$,

\bea
L=\frac{1}{4}(\dot{z}_1^2-\dot{z}^2_2)+\frac{\gamma}{2}(z_2\dot{z}_1-z_1\dot{z}_2)+\frac{\omega^2}{2}z^2_2-\frac{g}{2z^2_2}.
\label{lag}
\eea

\noindent This Lagrangian describes a coupled oscillators having balanced loss and gain and are interacting
with each other via long range inverse square potential and a two-body harmonic term. 
The Eqs. of motion can be obtained as
\bea
\ddot{z}_{1}+2 \gamma\dot{z}_{2}=0,
\label{nce1}\\
\ddot{z}_{2}+2\omega^2z_2+2 \gamma\dot{z}_{1}+\frac{2g}{z^3_2}=0.
\label{nce2}
\eea
The above Eqs. can also be derived from the following Hamiltonian

\bea
H=(P^2_{z_1}-P^2_{z_2})-\gamma(z_1P_{z_2}+z_2P_{z_1})-\frac{\omega^2}{2}z^2_2-\frac{\gamma^2}{4}(z^2_1-z^2_2)+\frac{g}{2z^2_2},
\label{classicalH}
\eea
where $P_{z_1}$ and $P_{z_2}$ are respectively the momenta conjugate to the normal coordinates $z_1$ and $z_2$.
\noindent Eqs. (\ref{nce2}) will be decoupled in terms of $z_1$ and $z_2$ as $\gamma$ tends to zero with
$z_1$ describing a free particle and $z_2$ describing a harmonic oscillator in a inverse
square potential. Integrating Eq. (\ref{nce1}), we get 
\bea
\Pi=\dot{z}_1+2\gamma z_2=2P_{z_1}+\gamma z_2
\label{z1}
\eea
with $\Pi$ being a constant of
integration which can be determined by fixing the initial conditions. It may be 
noted that $\Pi$ is an integral of motions related to the translational symmetry of the action.
The Poisson bracket of $\Pi$ with the Hamiltonian $H$ is zero. 
Thus the existence of two  integral of motions
$H$ and $\Pi$ in involution imply that the system is integrable.

\noindent Substituting Eq. (\ref{z1}) in (\ref{nce2}), we get
\bea
\ddot{z}_{2}+\Omega^2z_2+\frac{2g}{z^3_2}=-2 \gamma \Pi,\  \  \Omega^2=2(\omega^2-2\gamma^2).
\label{z2}
\eea

\noindent The frequency $\Omega$ is real for the range $-
\frac{\omega} {\sqrt{2}} \le \gamma \le \frac{\omega} {\sqrt{2}}$. This indicates 
$\cal{PT}$-symmetric phase transitions one at $-\frac{\omega} {\sqrt{2}}$ and other at $\frac{\omega} {\sqrt{2}}$. 
It is interesting to note that
in \cite{ben} the phase transition point depends on the linear coupling strength $\epsilon$ but in the present case
the phase transition point does not depend directly on the coupling parameter $g$ rather, as we shall see, the solution
of $z_1$ and $z_2$ put a restriction on the possible range of $g$ . 
 If we chose the initial
 conditions as to set $\Pi=0$, then Eq. (\ref{z2}) reduces to the Ermakov-Pinney equation of the 
following form:
\bea
\ddot{z}_{2}+\Omega^2z_2+\frac{2g}{z^3_2}=0.
\label{ep}
\eea  
This equation describes the motion of a particle in a harmonic plus inverse square
interaction. The system admits stable solutions for attractive inverse
square potential implying that $g<0$.
One of the possible ways to make $\Pi=0$, is to take the ratio
$\frac{\dot{z}_1(0)}{z_2(0)}=-2\gamma $. One such choice is
$\dot{z}_1(0)=-2\gamma b$, $z_2(0)=b$ and $\dot{z}_2(0)=a$. In this case the solution
of (\ref{ep}) is given as:

\bea
z_2(t)=\left [\frac{1}{b^2\Omega^2}\left (a^2 b^2-2g\right )\sin^2{\Omega t}+\frac{2ab}{\Omega}\sin{\Omega t}\cos{\Omega t}+b^2\cos^2{\Omega t}\right ]^{\frac{1}{2}}.
\label{epz2}
\eea

\noindent The expression for $z_1$ can be obtained by integrating Eq. (\ref{z1}),
\bea
z_1=-2\gamma\int{\left [\frac{1}{b^2\Omega^2}\left (a^2b^2-2g\right )\sin^2{\Omega t}+\frac{2ab}{\Omega}\sin{\Omega t}\cos{\Omega t}+b^2\cos^2{\Omega t}\right ]^{\frac{1}{2}}}dt+I
\label{intz1}
\eea
with I being the constant of integration which can be fixed
by the value of $z_1(t)$ at time $ t=0$.
We make the following change of variable
\bea
\sin{\phi}=\sqrt{\frac{D+(A-C)\cos{2\Omega t}-B\sin{2\Omega t}}{2D}},
\eea
where
\bea
A&=&\frac{1}{b^2\Omega^2}(a^2b^2-2g), \ \ \ \ \ B=\frac{ab}{\Omega},\ \ \  C=b^2,\ \ \ 
D=\sqrt{(C-A)^2+B^2}.
\label{abcd}
\eea
In the variable $\phi$ the integration of Eq. (\ref{intz1}) takes the following form:
\bea
z_1(t)&=&2\gamma\frac{\sqrt{C+A+D}}{\sqrt{2}\Omega}\int^{\phi}_{0}(1-\frac{2D}{(C+A+D)}\sin^2{\phi})^{
\frac{1}{2}}d\phi+I\\
&=&2\gamma \frac{\sqrt{C+A+D}}
{\sqrt{2}\Omega }E[\phi, 
k^2]+I
\label{epz1}
\eea
with $ E(\phi,k^2)$ being the elliptical integral of second kind having the argument $\phi$ and $k^2=\frac{2D}{C+A+D}$. 
The argument $k^2$of the elliptical integral satisfies the condition $0 <\frac{2D}{C+A+D}<1$.
In the unbroken $\cal{PT}$- symmetric region A and B are always real implying that D is also
real in this region. In order that the argument of the elliptical function be real the following 
condition must be satisfied:
\bea
D+(A-C)\cos{2\Omega t}\ge B\sin{2\Omega t}.
\label{res}
\eea
The expression of D in (\ref{abcd}) implies that D is real and positive and
$D>B$ as well as $D>(C-A)$. The maximum value of the right hand side
of the inequality in (\ref{res}) is B but at that time the left hand side becomes
D. Again, the minimum value  of the left hand side
of the inequality in (\ref{res}) is $D\pm(C-A)$ depending on the relative
value of A and C but in this case the right hand side of (\ref{res}) is zero and we 
have $D\pm (C-A)\ge 0$ which is always true. Thus, we can infer that the relation in
(\ref{res}) is true for all $t$.\\


Some observations regarding the nature of the solutions are as follows:

\begin{itemize}


\item 

As $\gamma$ approaches the transition 
value $\pm\frac{\omega}{\sqrt{2}}$,  $\Omega \rightarrow 0$, and $z_2$
becomes singular indicating the occurrence of the phase transition.

\item   The stability of the system depends on the nature of the solutions 
for $z_1$ and $z_2$. The solution (\ref{epz2}) for $z_2$ is well behaved and does not
introduce any instability in the system. However, the nature of the solution for $z_1$ 
depends on the initial conditions imposed and may 
introduce instability in the system. For example, for $a=1,b=1$, the expression 
for $z_1$ can be obtained from Eq. (\ref{epz1}):
\bea
z_1=2\gamma \frac{\sqrt{1+A+D}}
{\sqrt{2}\Omega }E[\phi, 
k^2]
\label{int}
\eea
with $A, B, C, D$ are given by Eq. (\ref{abcd}) 
and the constant of integration is taken to be zero. This solution has a 
periodic nature and does not introduce instability in the system. 
However, for $a=0,b=1$, the expression for $z_1$,
\bea
z_1=2\gamma\frac{E[\Omega t, 1+\frac{2g}{\Omega^2}]}{\Omega}
\eea
with $A, B, C, D$ are given by Eq. (\ref{abcd}) and $I=0$, 
increases linearly with time and introduces instability in the system.
Thus, the nature of the solutions for $z_1$ and therefore the stability of the
system depends on the initial conditions imposed.

\subsection{Two body Sutherland model with balanced loss and gain}

\par The Lagrangian in Eq. (\ref{lag0}) for $\epsilon =0$ describes 
 two body Sutherland\cite{sut} model with balanced loss and gain.
In this case Eqs. (\ref{noreom})  take the following form:

\bea
\ddot{z}_{1}+\omega^2 z_1+2 \gamma\dot{z}_{2}=0,
\nonumber\\
\ddot{z}_{2}+\omega^2z_2+2 \gamma\dot{z}_{1}+\frac{2g}{z^3_2}=0.
\label{nce02}
\eea

\noindent In the limit $\gamma$ tends to zero the above two Eqs. decoupled in terms of $z_1$ and $z_2$, 
one gives a harmonic oscillator and other a harmonic oscillator in a inverse square potential. It is the sole
effect of the dissipative term that make the system a coupled and nontrivial one. Interestingly the coupling
arising due to the inverse square potential in $(x,y)$ coordinates disappears in ($z_1, z_2$) coordinates.\\

\noindent We define $\dot{z}_1=p$ and $\dot{z}_2=q$ in order to investigate the possible equilibrium points and the
stability of the system. The stationary points
are those for which the following equations are satisfied:

\bea
\dot{p}+\omega^2 z_1+2\gamma q=0,\nonumber\\
\dot{q}+\omega^2 z_2+2\gamma p+\frac{2g}{z^3_2}=0,\nonumber\\
\dot{z}_1=0,\  \  \  \  
\dot{z}_2=0.
\label{eqi}
\eea

\noindent If we solve the above Eqs., we get
$(\tilde{p},\tilde{q},\tilde{z_1},\tilde{z_2})=(0,0,0,(\frac{-2g}{\omega^2})^{\frac{1}{4}})$ as the equilibrium point
 of the system which indicates that the coupling strength $g$ must be negative.
 In order to do the linear stability analysis, we consider a small variation about
the equilibrium point $(p,q,z_1,z_2)=(\tilde{p}+v_1,\tilde{q}+v_2,\tilde{z_1}+v_3,\tilde{z_2}+v_4)$.
A Taylor series expansion of Eqs. (\ref{eqi}) about the equilibrium point then yields up to first order, the following
set of Eqs.

\bea
\bp
\dot{v}_1\\
\dot{v}_2\\
\dot{v}_3\\
\dot{v}_4
\ep
=
\bp
0 & -2\gamma &  -\omega^2 &  0\\
-2\gamma &0& 0 &-4\omega^2\\
1&0&0&0\\
0&1&0&0
\ep
\bp
v_1\\
v_2\\
v_3\\
v_4
\ep
\label{mat}
\eea 

\noindent The eigenvalues of the $4 \times 4$ matrix in Eq. (\ref{mat}), given by
\bea
\lambda=\pm [\frac{P\pm \sqrt{P^2-16\omega^4}}{2}]^{\frac{1}{2}},\ \ \ P=(5\omega^2-4\gamma^2),
\eea
 \noindent will determine the nature
of the equilibrium point. Since positive eigenvalues indicate a growing solution, for
stable solution we must have all the eigenvalues either be negative or imaginary.
The only possibility of having stable solution is to take
 $\gamma=\pm \sqrt{\frac{5}{4}}\omega$ which makes all the eigenvalues imaginary. Therefore,
for the stable solutions the value of $\gamma$ should be in the range $-\sqrt{\frac{5}{4}}\omega<\gamma< \sqrt{\frac{5}{4}}\omega$.

The Eqs. (\ref{nce02}) define a coupled set of second order nonlinear differential equations. 
Unlike the case of rational Calogero model, we are unable to find any exact solutions. This is
primarily due the fact that a constant of motion similar to $\Pi$ can not be found. We investigate the 
solutions by employing  Lindstedt-Poincare perturbation method. It may be noted in this regard that
the system becomes decoupled in the limit $\gamma \rightarrow 0$.
We treat the $\gamma$ dependent term as perturbation with $\gamma$ being a small parameter. The
unperturbed Eqs. have the following form:

\bea
\ddot{z}_{1}+\omega^2 z_1=0,
\nonumber\\
\ddot{z}_{2}+\omega^2z_2+\frac{2g}{z^3_2}=0.
\label{nce03}
\eea

We employ the following initial conditions:

\bea
z_1(0)=0.5,\ \ \ \ z_2(0)=1\\
\dot{z}_1(0)=0,\ \ \ \ \dot{z}_2(0)=0.
\eea

Up to first order in $\gamma$, the solutions may be written as:

\bea
z_1(t)&=&.5\cos{\omega t}+O(\gamma^2),\\
z_2(t)&=&\frac{-2g}{\omega^2}+\gamma [-\frac{.5}{3 \omega}\cos{2\omega t}\\
&+&\frac{.5}{3 \omega}\{2\sin^3{\omega t}\sin{2\omega t}
+(1+\sin{2\omega t})\cos{\omega t}\cos{2\omega t}\}]+O(\gamma^2)
\eea
The integrability of this system is obscure and we obtain numerical solutions.  This numerical
results are in well agreement with the perturbative results at the initial time (fig(\ref{fig3})).

\begin{figure}[h]
\begin{center}
\fbox{\includegraphics[width=6cm,height=3 cm]{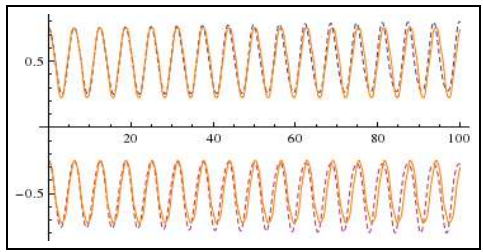}}
  \caption{(Color online) Numerical result (dashed line) vs perturbative result (continuous line). 
In this figure, $\gamma=.1$, $g=-0.5$, $\omega=1.0$. The upper graph
shows variation of $x$ as a function of time and the lower one shows variation of
y as a function of time.}
  \label{fig3}
\end{center}
\end{figure}

\end{itemize}

\section{Quantum case}

In order to quantize the classical Hamiltonian (\ref{classicalH}), we replace the classical variables
$P_{z_1}, P_{z_2}, z_1, z_2$ by operators satisfying the commutation relations $[z_1,P_{z_1}]=i, [z_2, P_{z_2}]=i$ and the rest 
of the commutators are zero. We replace $P_{z_1}$ by $-i\partial_{z_1}$ and $P_{z_2}$ by $-i \partial_{z_2}$ and 
obtain the following expressions for the Hamiltonian $\hat{H}$ and the conserved quantity $\hat{\Pi}$ in the quantum theory:
\bea
\hat{H}&=&\ [-(\partial^2_{z_1}-\partial^2_{z_2})+i \gamma(z_1\partial_{z_2}+
z_2\partial_{z_1})-\frac{\omega^2}{2}z_{2}^2+\frac{g}{2z_{2}^2}-
\frac{\gamma^2}{4}(z^2_{1}-z^2_{2})\ ],\\
\hat{\Pi}&=&  (-2i\frac{\partial}{\partial z_1}+\gamma z_2).
\eea 
The operators $\hat{H}$ and $\hat{\Pi}$ commute and constitute two integrals of motion for the system
which in turn implies that the system is integrable. Since $\hat{H}$ and $\hat{\Pi}$ commute, it is always possible
to chose a basis in which simultaneous eigen states of  $\hat{H}$ and $\hat{\Pi}$ may be constructed.
 An eigen function of the operator $\hat{\Pi}$ with continuous eigen
values $k$ has the following form:
\bea
\psi=\exp[\frac{i}{2}z_1(k-\gamma z_2)]\tilde{\phi}(z_2),
\label{eigpi}
\eea
where $\tilde{\phi}(z_2)$ is an arbitrary function of $z_2$ whose functional form is to be 
determined by demanding that $\psi$ is also an eigen function of the Hamiltonian H. 
In the limit of vanishing $\gamma$, the $z_1$ dependent part of $\psi$ is the
  wave function of a free particle with wave vector $\frac{k}{2}$. This is consistent with the fact that in the 
same limit $\Pi$ reduces to conjugate momentum operator $P_{z_1}$ up to an overall multiplication factor of two.
Substituting (\ref{eigpi}) in the time independent Schrodinger equation,
\bea
\ [-(\partial^2_{z_1}-\partial^2_{z_2})+i \gamma(z_1\partial_{z_2}+
z_2\partial_{z_1})-\frac{\omega^2}{2}z_{2}^2+\frac{g}{2z_{2}^2}-
\frac{\gamma^2}{4}(z^2_{1}-z^2_{2})\ ]\psi=E\psi,
\label{tisc}
\eea
we get the following equation:
\bea
\partial^2_{z_2}\tilde{\phi}(z_2)-\frac{1}{4}\Omega^2 z_2^2\tilde{\phi}(z_2)+\frac{g}{2z_2^2}
\tilde{\phi}(z_2)+\frac{k}{4}(k-2\gamma z_2)\tilde{\phi}(z_2)=E\tilde{\phi}(z_2).
\label{iso}
\eea
This is a differential equation of only one variable $z_2$. For vanishing $\gamma$, the centre of mass
and the relative coordinates separate out in the Hamiltonian (\ref{classicalH}) and the 
Eq. (\ref{iso}) describes the motion of an isotonic oscillator. However, for $\gamma \ne 0$, the centre of mass
modes are coupled to the equation governed by $z_2$. Similar situation arises in case $\epsilon =-\omega^2$ for
the system considered in Ref.\cite{ben} for which the Hamiltonian separates out into a free particle in centre of mass frame and a
simple harmonic oscillator in the relative coordinate for $\gamma=0$. The term linear in $k$ in Eq. (\ref{iso})
for $\gamma \ne 0$ can always be absorbed by a shift of the relative coordinate for harmonic oscillator,
but, not for isotonic oscillator. This poses difficulty in solving the Eq. (\ref{iso}).

The series solution method allows normalizable solution of (\ref{iso}) only for $k=0$. Other nontrivial exact 
solutions are also not apparent. This is consistent with the fact that we obtain
exact classical solutions when  the value of the constant of motion $\Pi$ is zero. Therefore, an obvious choice
is to consider the exact normalizable solution corresponding to $k=0$. In this case the Eq. (\ref{iso}) reduces
to the following form:
\bea
\partial^2_{z_2}\tilde{\phi}(z_2)-\frac{1}{4}\Omega^2 z_2^2\tilde{\phi}(z_2)+\frac{g}{2z_2^2}
\tilde{\phi}(z_2)=E\tilde{\phi}(z_2),
\label{isok0}
\eea
This equation is invariant under the operation $z_2\rightarrow -z_2$. Therefore the solutions
may always be chosen to be either even or odd under parity transformation.
 The potential has a singularity at $z_2=0$ which breaks the
space into two disjoint regions $(z_2>0$ or $z_2<0)$ and the wave function vanishes at $z_2=0$. 
The ground state wave function and the corresponding energy of (\ref{isok0}) are respectively given as 
\bea
\tilde{\phi}_{0}(z_2)=z^{\lambda}_2 \exp[-C z_2^2],\ \ \ \ \  E_0=-(2+4\lambda)C 
\eea
with $C=\pm\frac{1}{4}\Omega$ and $\lambda$ satisfying the relation $\lambda(\lambda-1)=-\frac{g}{2}$.
It is interesting to note that $C$ is real for the range $-
\frac{\omega} {\sqrt{2}} \le \gamma \le \frac{\omega} {\sqrt{2}}$. This indicates 
$\cal{PT}$-symmetric phase transitions one at $-\frac{\omega} {\sqrt{2}}$ and other at $\frac{\omega} {\sqrt{2}}$. 
Outside this range the energy is complex and comes in complex conjugate pairs indicating a broken $\cal{PT}$-symmetric
region. Interestingly the classical and quantum $\cal{PT}$-symmetric phase transitions occur at the same value of 
the parameter $\gamma$. 
In order to have the complete spectra we make the following substitution:
\bea
\tilde{\phi}(z_2)=z^{\lambda}_2 \exp[-\frac{1}{4}\Omega z_2^2]\phi(z_2),
\eea
with which Eq. (\ref{isok0}) reduces to the following form:
\bea
\partial^2_{z_2}\phi+(\frac{2\lambda}{z_2}-4Cz_2)\partial_{z_2}\phi=(E+2C+4\lambda C)\phi,
\label{norphiz2}
\eea
\noindent 
We demand a series solution of Eq. (\ref{norphiz2}) having the form
\be
\phi(z_2)=\sum^{\infty}_{n=0}a_n z^n_2.
\label{ser}
\ee
 \noindent

\noindent The recursion relation for $a_n$ is given by
\bea
a_{n+2}=\frac{(E+2C+4\lambda C)+4Cn}{(n+2)[(n+1)+2 \lambda]}a_n
\eea
with $a_1=0$, and
$a_0$ is obtained from the normalization condition.
 Thus the series for $\phi(z_2)$ contains only even powers of $z_2$. 
 For normalizable solutions, the series must be terminated and we get the following
expression for the energy states
\be
E=-2C(2n+1+2\lambda),
\label{en}
\ee
with $E=-2C(1+2\lambda)$ being the ground state energy.  If we now
chose the positive sign of $C$, then the ground state wave function is normalizable on the real line but the system becomes unbounded
from below, i.e the system does not have a stable ground state. So we chose the negative sign for $C$ which makes the system
bounded from below but in this case the normalization of the wave function becomes crucial and we discuss it below.
If we write $\phi(z_2)=\phi_{2m}(z_2)$ with $m=0,1,2....$, then
first few polynomials may be written as
\bea
\phi_2=a_0(1-\frac{4C}{(1+2\lambda)}z^2_2),\\
\phi_4=a_0(1-\frac{8C}{(1+2\lambda)}z^2_2+\frac{16C^2}{(3+2\lambda)(1+2\lambda)}z^4_2).
\eea
All the eigen states of $\hat{H}$ of the form $\phi(z_2) \psi_0$ with $\psi_0=z_2^{\lambda}\exp[-\{C z^2_2+\frac{i\gamma}{2}z_1 z_2\}]$,
are the eigen states of $\hat{\Pi}$ belonging to the zero eigen value.

We now consider the normalization of the ground state wave function

\bea
\psi_0=z_2^{\lambda}\exp[\frac{\Omega}{4}z_2^2-\frac{i\gamma}{2}z_1z_2],\ \ \ \ C=-\frac{1}{4}\Omega.
\eea

\noindent Clearly this wave function is not normalizable along the real $z_2$ line. We have to fix the Stoke wedges
in the complex $z_2$ plane where the wave function $\psi_0$ is normalizable. The first part ($z^2_2$-part)  of the exponential
vanishes in a pair of stoke wedges with opening angle $\frac{\pi}{2}$ and centered about the positive and
negative imaginary axes in the complex $z_2$-plane. The second part ($\frac{i\gamma}{2}z_1z_2$) of the exponential 
vanishes in the upper half of the complex $z_2$-plane if the coefficient  ($\frac{\gamma}{2}z_1$) of $z_2$ is negative and vanishes
in the lower half of the complex $z_2$ plane if the coefficient  ($\frac{\gamma}{2}z_1$) of $z_2$ is positive. 
Therefore $\psi_0$ vanishes  in a single stoke wedges either with opening angle $\frac{\pi}{2}$ and centered about 
the positive imaginary axes in the complex $z_2$-plane or with opening angle $\frac{\pi}{2}$ and centered about the 
negative imaginary axes in the complex $z_2$-plane.

\begin{figure}[h!]
\begin{center}
\fbox{\includegraphics[width=6cm,height=5 cm, ]{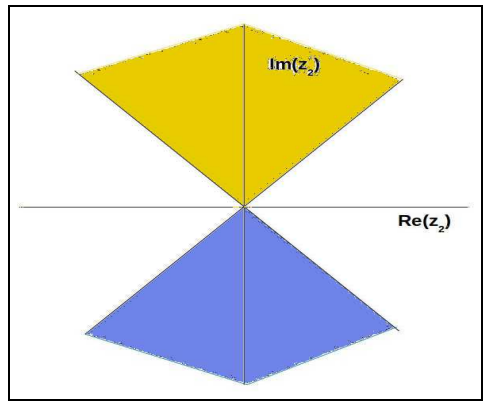}}
  \caption{(Color online): Stoke wedges. For negative coefficient of $z_2$, the Stoke wedge is centered about
the positive imaginary axis with a opening angle $\frac{\pi}{2}$. For positive coefficient of $z_2$, the Stoke wedge is centered about
the negative imaginary axis having the same opening angle.}
  \label{omega}
\end{center}
\end{figure}

\section{Conclusion}

We have considered a two body rational Calogero model
 having balanced loss and gain. 
The Hamiltonian for the system is obtained  which is found to be $\cal{PT}$ symmetric. 
This system admits two integral of motions in involution. This system is integrable both classically and
quantum mechanically.  
In particular, the classical Eqs. of motion for the system are solved exactly for the particular
 ranges of the parameters. We obtained exact, stable classical solutions.
We also quantized this classical model. The quantized system yields bound state solutions for exactly the 
same range of the parameters  for which the classical solutions are stable. The normalization of the wave functions in 
the proper Stoke wedges are discussed.
Further, the Calogero model with balanced loss and gain
 is studied classically, when the pair-wise harmonic interaction term is replaced by a
common confining harmonic potential. This system may be considered as 
the Sutherland model in the presence of balanced loss and gain. The integrability
of this system is obscure. In the classical level,
the stability analysis is carried out and perturbative solutions are obtained. 
Finally, this perturbative results are compared with the  numerical results.
In our study we only focus on two-body problem. The question of many-body generalization 
of coupled oscillators system having balanced loss and gain and are interacting via Calogero-Sutherland type 
of potential will be very much interesting.

\section{Acknowledgements}
This work is partly supported by a grant({\bf DST Ref. No. SR/S2/HEP-24/2012})
from Science \& Engineering Research Board(SERB), Department of Science
\& Technology(DST), Govt. of India. {\bf DS} acknowledges a research
fellowship from CSIR.

\end{document}